# FROM TRACES TO MEASURES: A PSYCHOMETRIC APPROACH TO USING LARGE LANGUAGE MODELS TO MEASURE PSYCHOLOGICAL CONSTRUCTS FROM TEXT

Joseph J.P. Simons[1], Liang Ze Wong [1], Prasanta Bhattacharya[1], Siyuan B. Loh[1], Wei Gao[2]

[1] Institute of High Performance Computing (IHPC), Agency for Science, Technology and Research (A*STAR), 1 Fusionopolis Way, #16-16 Connexis, Singapore 138632, Republic of Singapore

[2] School of Computing and Information Systems, Singapore Management University, 80 Stamford Road, Singapore 178902

*Large language models are increasingly being used to label or rate psychological features in text data. This approach helps address one of the limiting factors of digital trace data – their lack of an inherent target of measurement. However, this approach is also a form of psychological measurement (using observable variables to quantify a hypothetical latent construct). As such, these ratings are subject to the same psychometric considerations of reliability and validity as more standard psychological measures. Here we present a workflow for developing and evaluating large language model based measures of psychological features which incorporate these considerations. We also provide an example, attempting to measure the previously-established constructs of attitude certainty, importance and moralization from text. Using a pool of prompts adapted from existing measurement instruments, we find they have good levels of internal consistency but only partially meet validity criteria.*

Recent developments in large language models (LLMs) provide new possibilities in extracting psychological information from text. These flexible models perform well at a range of natural language processing (NLP) tasks without the need for task-specific re-training (Brown et al., 2020; Qin et al., 2023). As such, they are increasingly being used for text processing, including not just popular NLP tasks (e.g., stance and sentiment detection) but also more psychological task such as inferring personality (Yang et al., 2023) and moral values (Roy et al., 2022). This usage of LLMs to assess deeper latent variables about users can help mitigate measurement issues faced by digital trace data - a key challenge for computational social science research (Lazer et al., 2021). Unlike data from instruments designed for scientific data collection, digital trace data (such as social media posts) are not collected for research purposes. Rather, they are stored as a side-effect of the functioning of digital platforms. As a result, the question of what underlying psychological processes they track remains unclear (Lazer et al., 2021; Wagner et al., 2020), leading to issues of inconsistent measures, excessive analytic flexibility, and a focus on operationalizations rather than underlying processes. An appropriately-prompted LLM may be able to mitigate these issues for text data by quantifying theoretically-relevant variables expressed through text. In essence, this would allow researchers to overcome the lack of an inherent target of measurement in the data by providing their own.

This task is a form of psychological measurement, and so should be guided by psychometric principles. Unlike surface-level text features (such as number of pronouns), psychological constructs (variables such as the author's attitudes, values, or personality) are latent. We observe them indirectly via their effects on downstream behaviours - in this case, linguistic expression. In this regard, LLM ratings of psychological variables are analogous to traditional self-report items on a questionnaire. For such applications, established psychometric procedures should be used to guide the prompting and assessment of these models. To this end, we offer three practical methodological recommendations. First, constructs should be assessed using multiple prompts rather than a single prompt. Researchers can draw on existing self-report scales to generate a pool of items. As each of these is a noisy indicator, the aggregate response is likely to be better than any one individual prompt. Second, in assessing model performance, it is important to first establish whether any latent construct is being captured at all. This can be verified via tests of reliability, including those that assess factor structure and internal consistency. Finally, while evaluation metrics (such as human ratings) have an important role in model assessment, they are correlates of the desired target of measurement and not a direct ground-truth measure. While we would want our models to agree with them to a degree, we should not be aiming to maximize these correlations as the evaluation metrics are themselves error-prone indicators of the desired target of measurement.

**Digital trace data and measurement**

With ongoing increases in digitalization, more and more human activity is being captured in large digital datasets which hold great potential for understanding human behaviour (Conte et al., 2012; Kitchin 2014; Lazer et al., 2009). A wide range of everyday behaviours now take place online or via other digital platforms, creating lasting digital logs. These digital traces offer a naturalistic record of human behavior in the wild and are often very granular, providing an unprecedented opportunity to study behaviour at scale. Indeed, the analysis of trace data has provided novel insight into topics as diverse as the spread of misinformation over online platforms (Bakshy et al., 2015; Vosoughi et al., 2018), the role of social factors in health process (Bargain and Aminjonov, 2020; Hobbs et al., 2016) and the process of scientific knowledge creation (Li et al. 2019, Park et al., 2023).

However, and as mentioned above, this trace data is not generally gathered for scientific purposes. As such, it does not usually have an inherent target of measurement (Lazer et al., 2021). Purpose-built scientific measurement instruments are designed to assess a specific and pre-determined target construct. For instance, a personality inventory is designed, built, and evaluated as a way to reliably measure personality. Trace data, conversely, are often generated as a by-product of digital platforms. As such, what behavioural processes they are indexing is often not clear. We know what a thermometer reading means because thermometers were built to measure temperature, but it is far less clear what a Twitter hashtag measures, or if it measures anything reliably at all (e.g. Daer, 2014).

This has at least three potential negative consequences. First, an inconsistent use of metrics across papers can pose challenges. A given metric may capture a range of different meanings - for example, a Retweet can indicate endorsement or dislike depending on the context of its use (Tufecki, 2014). As a result, different researchers end up using the same metric to indicate different phenomena (e.g., Freelon, 2014), an example of the jingle fallacy (Kelley, 1927). This cross-study variability in operationalization leads to difficulties in interpreting results (Lazar et al., 2021). Second, poor measurement undermines confidence in the conclusions of a study (Flake & Fried, 2020; Flake et al., 2017). Low methodological consistency in measurement, such as having multiple operationalizations of the same construct available, provides researchers with an additional degree of flexibility in modelling. This increases the probability of mistaking sampling noise as a genuine effect (Simmons et al., 2011). Third, there is a risk of making traces themselves the objects of study. Like the person searching for their keys under the lamppost because that's where the light is, we end up substituting the phenomena which is easily quantified for the one we are actually interested in. Lazer et al. (2021) contrast the study of Twitter users with the study of Twitter accounts; while they make this point about the issue of bot accounts, the contrast expresses the more general risk well. Much as Baumeister et al. (2007) argue psychology has lost track of actual behaviour and become "the science of self-reports and finger pushes", we may worry about fields like NLP and computational social science losing track of people and becoming a science of Retweets and Follows.

**Applying large language models to the task of psychological measurement**

Researchers are increasingly turning to LLMs as a fast and cost-effective approach to labelling text. Due to their flexibility, these models provide researchers with versatile tools for converting unstructured textual data into quantitative scores and labels on desired features (e.g. Ziems, 2024). This has led to rapid LLM adoption in popular tasks such as sentiment analysis and stance detection (e.g., Zhang et al., 2023; Cruickshank and Ng, 2023). Increasingly, LLM use is being extended to user-level topics, such as author personality inference (Yang et al., 2023) and moral value detection (Roy et al., 2022), which had remained understudied. LLMs are extremely well-positioned to perform these tasks due to two key characteristics: (a) being able to respond to in-context meaning and (b) being able to attempt novel tasks without custom model development, re-training, or fine-tuning (Demszky et al., 2023).

When researchers prompt a model to assess text features such as opinions, values, or author personality, they are engaging in psychological measurement. The resulting ratings are intended to provide observable but noisy indicators of underlying latent variables. The models are not being asked to enumerate surface-level textual features (e.g., review sentiment) about a post or message, but provide an estimate of deeper, underlying factors about the author of the content. LLM ratings can therefore be thought of in much the same way a personality questionnaire, which provides observable indicators of latent personality dimensions. To be clear, we do not intend to claim that LLMs complete these measurement tasks in a human-like way, just that their responses have a similar role to human ratings as observable indicators of a latent factor.

This psychometric conceptualization makes clear how LLMs can address the measurement issues in text-based trace data. By providing a method to quantify latent variables as expressed in text, LLMs can enable text-based trace data to be connected to the theoretical constructs of social-personality psychology; data such as posts and replies can be quantified in the theoretical vocabulary of attitudes, persuasion, intergroup processes, and individual differences. Without LLMs, this would require an unfeasibly large amount of human rating labour to achieve at scale. Such an approach would provide computational social science with stronger measurement models, and social-personality psychology with a novel naturalistic data source – a win-win situation.

## A psychometric approach to using large language models to measure psychological constructs

Psychology and allied disciplines have developed a number of relevant techniques for assessing how well psychological measures perform. Importantly, these are based on the assumption that the constructs being measured are inherently unobservable and hence not amenable to direct ground-truth measurement (DeVellis & Thorpe, 2021; Kaplan & Saccuzzo, 2001; see Flake et al., 2017 for a review of these in practice). These techniques center around the twin concepts of *reliability* (how well the measure is assessing any underlying construct; Revelle & Condon, 2018) and *validity* (whether the measure is assessing the desired underlying construct; Cronbach & Meehl, 1955). The former is often assessed by the internal coherence of different measures of the same construct, the latter by whether measures correlate with expected associates of the underlying construct. These can be seen as sequential considerations - before determining if we are measuring what we intended to measure, we first need to ensure we are measuring something.

These same techniques can be applied when assessing psychological constructs using LLMs. While self-report questionnaire items are more common in psychology, the basic principle is the same. In both cases, we generate observable data (i.e., self-report ratings or LLM responses) which are intended to measure an unobservable latent variable (i.e., a psychological construct, such as personality). The challenge then is to establish how well they are doing so.

Taking a psychometric approach to LLM-based measurement suggests three key methodological take-aways for the use of language models in this domain.

First, there is value in assessing each desired construct with multiple prompts. An average of responses to prompt variants is likely to work better than any single one. Each rating is a noisy indicator of an underlying latent construct; as such, combining the results of multiple prompts will give a better answer than any one in isolation. Fortunately, existing scales provide a fertile source of prompt variants. Many psychological constructs have one or more associated multi-item rating scales which have been systematically developed and validated. As LLMs can respond to free-text prompting, these items can be adapted to serve as a pool of instructions for assessing the construct.

Second, in evaluating model performance, it is important to account for internal consistency prior to assessing correlations with external outcomes. Current evaluation procedures focus on the correlation of model results with external criteria, such as human ratings. These correlations are an important part of measure validity (see point 3 below). However, this addresses the question of *what* is being measured without considering the issue of whether *anything* is being measured. Providing an initial assessment of reliability is not just epistemically virtuous, it also provides direct benefits to researchers. Measurement unreliability attenuates correlations with other factors, and so an unreliable measure is unlikely to predict a ground-truth criterion, even when the underlying latent variable is a good predictor. Identifying low reliability early can thus avoid wasted effort on trying to validate a poorly-functioning measure. There are a range of quantitative techniques for assessing this along with open-source software implementations (Jorgensen et al., 2022; Revelle, 2023; Rosseel 2012). Importantly, these are internal tests – they do not require any data other than the measures themselves, and so are not burdensome to implement.

Finally, while validating model ratings against external evaluation variables is important, these should be taken as validation criteria rather than ground-truth measures. To be confident LLM ratings are measuring what they are intended to measure, it needs to be shown that they predict expected outcome variables. However, these validation criteria are not ground-truth measures. By definition, this target is latent and so not amenable to direct quantification. Rather, these criteria represent additional likely correlates of the construct we wish to measure (its nomological net; Cronbach & Meehl, 1955). To give an example, we would be suspicious of a measure of extraversion which does not predict peer ratings of extraversion. However, this discrepancy is a difference between two indicators of the same underlying construct, not a failure of the model to recreate the "true" values. By the same token, we would not want to optimize how well the model recreates the peer ratings – the aim is to create a separate measure of the same construct, not re-create an existing measure which may have its own biases.

### A proposed psychometric workflow for LLM-based measurement

Based on the considerations above, we proposed the following workflow for assessing psychological constructs with LLMs. This process is derived from similar approaches for self-report measures (e.g., Flake et al., 2017), and shown graphically in Fig 1.

1. **Identify target of measurement.** The first stage is to conceptually clarify what it is that needs to be measured. What precisely is the nature of the hypothesized latent variable and what should it correlate with? To this end, literature searches and tools such as the Semantic Scale Network (Rosenbush et al., 2020) can be used to identify relevant psychological constructs (and associated measures)

2. **Generate rating prompt pool.** To assess internal consistency, a pool of variant rating prompts is required. These should be alternate ways of asking the same basic question. Existing self-report instruments from psychology can provide a relevant pool of items for use in these rating prompts. These instruments have often gone through extensive selection from larger initial item pools, and so provide an empirically-grounded selection of items to assess the desired construct. Furthermore, the multi-option Likert response scales used in these measures provide the model with more granular response options than dichotomous classification tasks; more fine-grained responses provide better estimation of the associations between variables.

3. **Generate scores for a sample.** The resulting rating prompts can then be used to generate ratings for a sample of stimuli. As revisions to the item pool may be needed, it is prudent to then divide the data into three folds; one for exploratory tests of factor structure, one for assessing degree of internal consistency, and one for testing associations with validation criteria. This enables revisions to be made based on the initial results while limiting the possibility for over-fitting to the sample. One advantage of digital datasets is they are often quite large, enabling this kind of subdivision while still providing decent sample size for each fold.

4. **Reliability I: Assess factor structure from the ratings.** The variant prompts will have a certain implied factor structure. In the simplest case, they are unidimensional (i.e., all responses should load on a single latent factor). However, it may also be the case that different items are intended to assess different underlying variables. As with self-report items, the first issue to be addressed empirically is whether the expected factor structure emerges - are there the expected number of factors, and do the items load as expected? This can be assessed through standard techniques such as Exploratory Factor Analysis (EFA) to examine what factor structure emerges from the data, or Confirmatory Factor Analysis (CFA) to assess how consistent the data is with a hypothesized factor model. Based on these analyses, the item pool can be amended.

5. **Reliability II: Quantify internal consistency of each factor.** Once the factor structure has been confirmed, the internal consistency of each factor can be assessed. Historically, internal consistency has often been quantified with Cronbach's alpha. This metric is simple but makes strong assumptions (i.e., unidimensionality, equal loadings, uncorrelated residuals; e.g., Yang & Green, 2011). McDonald's omega (McDonald, 1999) is an alternative model-based reliability estimate which makes less strong assumptions (see Flora, 2020; Revelle & Condon, 2019 for tutorials).

6. **Validity: Assess correlations with external factors.** While the steps above provide evidence the prompts are measuring a latent construct, the nature of this construct remains untested. To address this, the correlation of aggregated LLM ratings with external criteria can be measured. So, for example, a measure of extraversion may be validated by examining its correlations with the size of online network. While human annotation or linking text to survey data are extremely useful for this purpose, digital trace text offers rich alternative possibilities for validation criteria. As it is generated in naturalistic contexts and often includes associated metadata, there are often candidate criteria such as text source (e.g., different subreddits), other text features (such as sentiment in the message or replies to it), or platform-specific features (such as Facebook's 'likes' and other affective 'reactions'). As a result, the more burdensome approaches of annotating digital logs or linking them to survey data are not always necessary for validation.

7. **Repeat over new samples and validation targets.** Validation is not a one-time evaluation. The validity of a measure will vary across data sources and populations. Thus, there is a need for ongoing process of validation, whereby the procedures above are replicated across different cases and validation criteria.

## AN EXAMPLE: ATTITUDE PROPERTIES

As a demonstration of the steps above, we provide an example of our own attempts to use LLM prompts to assess attitude properties. We adapt existing self-report measures from the attitudes literature to generate a pool of rating prompts, apply these to an existing benchmark dataset (the SemEval 2016 Stance Detection from Tweets dataset; Mohammad et al., 2016), assess the factor structure of the resulting scores, quantify the reliability of the resulting factor scores, and then check their validity against initial validation criteria.

### The target of measurement: Attitude certainty, importance and moralization

Attitudes are valenced evaluations of a target object (e.g., Eagley & Chaiken, 1993). For example, 'liking' (a positive evaluation) chocolate ice cream (the target object) is a positively-valenced attitude. Thinking the death penalty is morally objectionable is a negative attitude.

Attitudes differ beyond just their valence, however. Not all positive and all negative attitudes function in the same way. For example, a personal distaste for anchovies and a deeply-held opposition to capital punishment are importantly different, even though they are both negative attitudes. In terms of characterizing these differences, researchers have identified and empirically validated a range of attitude properties (Petty & Krosnick, 1995; Luttrell & Sawicki, 2020; Visser et al., 2006). Here we focus on three: *attitude certainty* (how confidently held the opinion is; e.g., Petrocelli et al., 2007), *attitude importance* (how much the topic matters to the attitude holder; e.g., Howe & Krosnick, 2017) and *moral conviction* (how much the attitude is connected to deeper moral values; e.g., Skitka, 2010).

We chose these characteristics as they are both individually and socially consequential. On the one hand, they are contributors to individual-level attitude strength (Luttrell & Sawicki, 2020; Visser et al., 2006). Strong attitudes are more stable over time, resistant to persuasion, and influential on thought and behaviour (Petty & Krosnick, 1995). Certainty, importance and moral conviction all contribute to how strong a given attitude is, although there may be situational moderators of this relationship, particularly for certainty (e.g., Clarkson et al., 2008). However, beyond individual-level attitude strength, these three properties have also been linked to various forms of intolerance (e.g., Skitka, Bauman & Sargis, 2005; Rios, DeMarree & Statzer, 2014; Zuwerink & Devine, 1996).

For this reason, being able to detect these properties in digital trace data would be beneficial for a range of applications. Examples include better ability to understand persuasion processes in digital contexts, enriched opinion detection beyond simple attribution of pro/anti stance, and early detection of potential social faultlines forming around particular issues.

**Generating alternative rating prompts**

To assess each of these attitude attributes, we need a pool of rating prompts, providing varying ways of asking an LLM to rate a given text on the same feature. A starting point is provided by existing validated self-report scales from empirical psychology. For attitude certainty, we used the validated and widely-used scale developed by Petrocelli et al. (2007). This seven-item instrument assesses separate factors of clarity (i.e., knowing what one's own opinion is) and correctness (i.e., thinking others should think the same as you). For attitude importance, we used two scales that are widely cited in political psychology (Boniger et al., 1995; Zuwerink & Devine, 1996). Each of these was 4 items long. We used two sets as they had not been subject to the same level of psychometric testing as the certainty scale. This allowed us to obtain a reasonably large pool of items such that some could be discarded if necessary. Finally, moral conviction has been previously measured in a number of ways; we used the four-item scale recommended by one of the leading researchers in the area on her website (Skitka, n.d.)

This initial pool of self-report items required some adaptation to be useable as response items in a prompt. First, they were worded in the first person and, in some cases, made reference to specific issue topics (e.g., capital punishment). For current purposes, all items were re-worded to ask about the attitude of the author of a text towards the issue discussed in the text. Second, different items use different response scales . To address this, all items were reworded to be statements, such that a common agree/disagree response scale could be used for all. For example, the Petrocelli et al. (2007) certainty scale has an item "How certain are you that your attitude toward capital punishment is the correct attitude to have?". Respondents provide a rating of their degree of certainty on a 1-9 scale, with 1 anchored as "not certain at all" and 9 as "very certain". We adapted this to "The author is certain that their attitude toward this issue is the correct attitude to have", which can be responded to with levels of agreement ranging from "strongly disagree" to "strongly agree". The final response items are given in Table 1.

**The dataset: Detecting Stance in Tweets (SemEval-2016)**

As a data source, we used the benchmark SemEval 2016 stance detection dataset (Mohammad et al., 2016). This public dataset consists of Tweets relevant to five targets (Atheism, Climate Change is a Concern, Feminist Movement, Hillary Clinton, Legalization of Abortion), with human annotations of author stance (i.e., whether the author favours, is against, or is neutral to the target). We chose this dataset due to its wide use in NLP and computational social science tasks, as well as the fact it contains existing human labels of stance . We combined the original training and test datasets to create a full dataset of 4,063 Tweets.

We divided this dataset into three approximately equal subsets; one for exploration of factor structure (exploratory), one for confirmatory testing of internal consistency (confirmatory), and one for testing correlations with validation criteria (validation). Doing so means we can update our measures based on the results at each step while also safeguarding against over-fitting to sample specific noise.

Using the prompts described earlier, we generated ratings for all 4,063 Tweets. With 19 rating items for each Tweet, this led to 77,197 total prompts generated using the previously developed template. These were then run through OpenAI's GPT-3.5 Instruct model (Ouyang et al., 2022), which is fine-tuned to follow instructions (as opposed to the GPT-3 or GPT-4 models which are optimized for dialogue). We used a temperature setting of 0 for lower variability across runs and hence greater reproducibility.[1] The total cost for these prompts was approximately $15 (USD).

In terms of data processing, responses were checked for consistency with the requested options and then transformed into numerical ratings. In our sample, there were a total of 5 cases where the LLM returned a response other than the requested options; these were treated as missing data. The text responses were then converted to numerical values (-2, -1, 0, 1, 2), and the negatively-coded items (Importance_2, Importance_3) reversed such that a high score always indicated a higher level of the construct in question.

**Assessing reliability I: The factor structure of the measures**

The first step was to examine the factor structure. The key question is whether the ratings recreate the expected factor structure; do we find distinct factors corresponding to certainty, importance and moralization. To this end, we used Exploratory Factor Analysis (EFA) to assess the structure of the LLM responses in an open-ended manner. This approach examines what factors emerge from the data, rather than comparing the relative fit of two or more candidate models. The analysis was conducted in R using the *psych* package (version 2.4.6.26; Revelle, 2023) on the exploratory subset of the data (n = 1,354 Tweets).

Initial factorization was conducted using the default minimum residual method. As the rating variables had only five possible values, the analysis was conducted on their polychoric (rather than Pearson) correlations; this avoids treating them as continuous.

To determine the appropriate number of factors to retain, we carried out parallel analysis (Horn,

[1] Setting at temperature of 0 reduces variability but still does not lead to deterministic behaviour, so responses might vary slightly across runs. However, responses would vary a lot less compared to setting the temperature to a larger number such as 0.7 or 1.

1965) to compare our factor eigenvalues against factor eigenvalues from random data of the same shape (*n* and *p*) as our data. This pointed to a five-factor solution; for the first five factors, the eigenvalues from the actual data are higher than those from the random data. As we expected the attitude properties to be correlated, we rotated the factors using the promax method which allows for correlated factors. The results obtained using oblimin rotation were similar.

Identifiable factors corresponding to the attitude properties emerged. The left-hand columns of Table 2 show the loadings; for ease of reading, values less than 0.3 in magnitude are suppressed. The items from a common set tended to load on the same underlying factor, leading to factors which can be identified as attitude certainty (F1), attitude importance (F2 + F4) and moral conviction (F3). The fifth factor did not have a clear meaning, seemingly just capturing additional residual correlation between items.

However, the loadings departed from a clear structure in three ways. First, the clarity and correctness items were not cleanly distinguished, with the first factor capturing variance across both sub-facets. Second, attitude importance was subdivided into two factors. This distinction was largely based on the difference between positively and negatively worded items rather than any meaningful difference in content, replicating a pattern identified in human respondents (who frequently struggle with reverse-coded items; e.g. Weijters & Baumgartner, 2012). Third, a number of items did not perform well; some loaded primarily on the wrong factor (Clarity_1, Correctness_1, Moral_2), whereas others showed substantial cross-loadings (Clarity_3, Morality_1).

As a result, we made the following modelling decisions. First, we only attempted to measure overall attitude certainty rather than separate clarity and correctness factors, as the distinction between these two did not emerge clearly in the current data. Second, we dropped the first three Importance items as these seemed to be loading on an entirely separate construct. Finally, we dropped the items which primarily loaded on incorrect factors (Clarity_1, Correctness_1, Moral_2); items which showed some cross-loading were retained if they exhibited their strongest association with the postulated construct (Clarity_3, Moral_1).

In summary , we propose that LLM ratings can measure attitude certainty (4 items; Clarity_1-3, Correctness_3), attitude importance (5 items; Importance 4-8) and moral conviction (3 items; Moral_1, Moral_3-4).

### Assessing reliability II: The internal consistency of the factors

With a candidate factor structure, we proceed to quantify the reliability of each measure in the unseen confirmatory dataset (n = 1,354 Tweets). To this end, we calculated both Cronbach's alpha and McDonald's omega. The former is more widely used but makes stronger assumptions; the latter is more complicated to calculate but also more flexible.

As omega is a model-based statistic, it is necessary to derive it from a factor model. As we had comparatively few indicators and an expected factor structure, we followed the more confirmatory approach provided by Flora (2020) (compared to a more exploratory approach; Revelle & Condon, 2019). To this end, we fitted confirmatory factor analysis models using R's laavan package (v. 0.6-18, Rosseel, 2012). For each set of ratings, we fitted a unifactor model with all indicators loading on a single latent variable. The relevant reliability statistics were then derived from these models using the semtools package (v. 0.5-6, Jorgensen et al., 2022). As with the EFA, we derived these models based on polychoric correlation matrices due to the small number of response options.

Despite being fitted to unseen data, the confirmatory factor models generally captured the proposed factors well. This indicates that the omega values are informative. The results shown in the right-hand columns of Table 2 highlight that all items loaded well on their related factors. Model fit statistics (i.e., robust CFI, robust TLI) were good for the certainty and importance factor; the conventional threshold is 0.95, and the lowest value for any model was 0.949. The robust RMSEA values failed to achieve the standard threshold for good fit (<= 0.6). However, RMSEA is known to be overly conservative in smaller models such as these (Kenny et al., 2015). Fit statistics for the moral conviction model are not reported as they were trivially perfect - as a three-indicator single-factor model, it can reproduce any pattern of correlations.

All three scales showed good levels of internal consistency. The key reliability indices are reported in the bottom-right of Table 2. As can be seen, values range from 0.82 to 0.97. Generally, a threshold of 0.7 is seen as acceptable and 0.8 for good, and so, by popular benchmarks, all three measures would be considered reliable.

**Assessing validity: Relationship with validation criteria**

Based on the tests above, we have evidence that the LLM prompts are measuring something about the texts; alternate prompts co-vary as expected and show a high degree of reliability. The remaining question is whether they are measuring the specific targeted constructs. As these constructs are not directly observable, this is not something which can be established in a conclusive manner. Rather, it remains an ongoing process of comparing the model scores to validation criteria.

Fortunately, the dataset has a useful validation criterion, the presence vs. absence of a stance. Multiple human raters classified each message in terms of the stance of the author towards the focal issue – either supportive, opposed, or no stance expressed. Taking a stance (e.g., supporting or opposing) is making a public commitment to a position (Du Bois, 2007; AlDayel & Magdy, 2021). Attitude strength is one likely determinant of whether a person undertakes such a public commitment. Attitudes that are more important and moralized tend to be more influential on behaviour (Howe & Krosnick, 2017; Skitka et al., 2021), and attitude certainty has been linked to attitudinal advocacy in particular (although with some nuances; Cheatham & Tormala, 2015, 2017). As such, we would expect stances to be associated with higher attitude strength and, as such, all three attitude properties to be higher for stance-expressive messages.

As an initial validity test of our prompts, we predicted the Tweet stance from the LLM-generated attitude property scores using logistic regression. This analysis was conducted in the unseen validation dataset ($n$ = 1,355). As there were three stance classes, we split this test into two tasks; (a) distinguishing supportive stances from non-stance Tweets and (b) distinguishing opposing stances from non-stance Tweets. Decomposing the task this way can accommodate valence effects (i.e., capture any asymmetric effects for supporting vs. opposing). Attitude property scores were calculated by averaging responses to the prompts selected in earlier phases. As the resulting scores showed a fairly high level of intercorrelation ($r$s > 0.62), we assessed their effects using separate models; estimating the unique effects of the three properties simultaneously would be challenging due to the high level of collinearity.

In addition to the base models predicting stance from attitude properties, we included two additional control measures.

*Sentiment control.* A critic may ask if our prompts are just assessing sentiment extremity, rather than any deeper property of the author's attitudes. Looking at Table 1, it is not implausible that a model would use emotionality as a cue for many of these responses. To address this possibility, each model was re-run with sentiment scores as a control. These were calculated using the VADER library for Python (Hutto & Gilbert, 2014); in this case, we used VADER's compound score as a measure of overall positive / negative sentiment. If the LLMs are just measuring sentiment, controlling for this should substantially reduce (if not eliminate) the relationship between attitude property scores and stance position.

*Comparison prompt.* We also compared the performance of our psychometric approach to a simpler comparison prompt. This was a single prompt based on the same template as in Fig. 2, but simply asked if the author felt strongly on the topic (with agree / disagree as the two response options). As such, it did not call on prior conceptual and measurement work, did not use multiple prompts, and had a less granular response scale.

The results are given in Table 3. The top section (a) shows models of supportive stances, and the bottom (b) for opposing stances. Each column gives the coefficients and model fit for a single model.

As can be seen in Table 3a, our psychometric prompts were reasonably good at distinguishing supportive stances from non-stance messages. The attitude properties were all significantly positive predictors of positive stance status. These effects were not strongly affected by adding sentiment scores to the models. The simpler comparison prompt did not perform as well; the relationship between scores and stance status was in the correct direction and significant, but the overall model fit was a lot lower.

However, the results were not as encouraging for opposing stances. Contrary to predictions, attitude property scores were lower for opposing stances than for non-stances (although this effect was only significant for attitude importance). The comparison prompt gave results in the correct direction and was significant, but the model fit (pseudo-R2) shows this was small.

**Discussion**

This illustration points to the value of our psychometrically-informed approach to using LLMs for textual rating tasks. By treating this as a psychological measurement task, we were able to clarify the latent constructs to be measured, develop behaviourally-informed prompts based on prior research, and assess the reliability of these prompts. The association with supportive stance expression shows the efficacy of this method, with the resulting scores out-performing a more straightforward single prompt strategy.

The mixed validation results also point to the need for careful evaluation of LLM measures. Our prompts are face valid and the resulting scores showed good reliability. However, taking this as evidence that they were good measures would have been premature as they did not show the expected associations with opposing stance expression. As we do not have ground truth measurements, this result is ambiguous. One possibility is that that our measures are not capturing the desired constructs. Rather, they measure something more like strength of positive evaluation. Another possibility is that negative stance expression is not a good validation criterion for attitude strength - maybe higher strength predicts advocacy for positive attitudes but not negative ones. Further work would be needed to disentangle these issues; as stated previously, validation is not a one-time exercise.

## CONCLUSION

In summary, we have outlined a procedure for designing, implementing and testing LLM prompts for extracting latent variables from text stimuli. This approach provides a novel conceptualization of LLM performance, as well as a structured process for developing and evaluating these measures. This method can also utilize existing resources from empirical psychology and conduct initial reliability assessment based purely on the LLM results, reducing the burden of applying LLMs to new rating tasks.

We see this approach as holding promise for addressing the measurement issues inherent in digital trace data such as social media text (Lazer et al., 2021; Wagner et al., 2021). Unlike with custom-designed measurement instruments, it is unclear what underlying phenomena these data are tracking. Our proposal is that LLMs have the potential to quantify the latent constructs manifested in text at scale, allowing unstructured text data to be characterized in terms of the latent constructs of social and personality psychology.

However, as our illustrative analysis shows, using LLMs for this purpose requires careful evaluation. Even if prompts are measuring some genuine property of the text, this may not be the desired construct. Furthermore, in the absence of ground-truth data, validation is not a straightforward process and may require iteration over a number of studies.

Our hope is that this paper can provide a useful and practical technique for researchers, promoting greater integration of the digital datasets of NLP and computational social science, with the theoretical resources of social and personality psychology.

**Funding acknowledgement.** This research is supported by A*STAR (C232918004, C232918005). This research is supported by the SMU-A*STAR Joint Lab in Social and Human-Centered Computing (SMU grant no.: SAJL-2022-CSS02, SAJL-2022-CSS003).

# References

AlDayel, A., & Magdy, W. (2021). Stance detection on social media: State of the art and trends. *Information Processing & Management, 58*, 102597

Bakshy, E., Messing, S., & Adamic, L. A. (2015). Exposure to ideologically diverse news and opinion on Facebook. *Science, 348*, 1130-1132.

Bargain, O., & Aminjonov, U. (2020). Trust and compliance to public health policies in times of COVID-19. *Journal of Public Economics, 192*, 104316.

Baumeister, R. F., Vohs, K. D., & Funder, D. C. (2007). Psychology as the science of self-reports and finger movements: Whatever happened to actual behavior? *Perspectives on psychological science*, *2*, 396-403.

Boninger, D. S., Krosnick, J. A., & Berent, M. K. (1995). Origins of attitude importance: self-interest, social identification, and value relevance. *Journal of Personality and Social Psychology*, *68*, 61.

Brown, T., Mann, B., Ryder, N., Subbiah, M., Kaplan, J. D., Dhariwal, P., ... & Amodei, D. (2020). Language models are few-shot learners. *Advances in Neural Information Processing Systems, 33*, 1877-1901.

Cheatham, L. B., & Tormala, Z. L. (2017). The curvilinear relationship between attitude certainty and attitudinal advocacy. *Personality and Social Psychology Bulletin, 43*, 3-16.

Cheatham, L., & Tormala, Z. L. (2015). Attitude certainty and attitudinal advocacy: The unique roles of clarity and correctness. *Personality and Social Psychology Bulletin, 41*, 1537-1550.

Chen, L., Zaharia, M., & Zou, J. (2023). How is ChatGPT's behavior changing over time?. *arXiv preprint arXiv:2307.09009*.

Clarkson, J. J., Tormala, Z. L., & Rucker, D. D. (2008). A new look at the consequences of attitude certainty: The amplification hypothesis. *Journal of Personality and Social Psychology, 95*(4), 810–825. https://doi.org/10.1037/a0013192

Conte, R., Gilbert, N., Bonelli, G., Cioffi-Revilla, C., Deffuant, G., Kertesz, J., ... & Helbing, D. (2012). Manifesto of computational social science. *The European Physical Journal Special Topics*, *214*, 325-346.

Cronbach, L. J., & Meehl, P. E. (1955). Construct validity in psychological tests. *Psychological bulletin, 52*, 281.

Cruickshank, I. J., & Ng, L. H. X. (2023). Use of Large Language Models for Stance Classification. *arXiv preprint arXiv:2309.13734*.

Daer, A. R., Hoffman, R., & Goodman, S. (2014, September). Rhetorical functions of hashtag forms across social media applications. In *Proceedings of the 32nd ACM International Conference on the Design of Communication CD-ROM* (pp. 1-3).

Demszky, D., Yang, D., Yeager, D. S., Bryan, C. J., Clapper, M., Chandhok, S., ... & Pennebaker, J. W. (2023). Using large language models in psychology. *Nature Reviews Psychology, 2*, 688-701.

DeVellis, R. F., & Thorpe, C. T. (2021). *Scale development: Theory and applications.* Sage publications.

Du Bois, J. W. (2007). The stance triangle. *Stancetaking in discourse: Subjectivity, evaluation, interaction, 164*, 139-182.

Eagly, A. H., & Chaiken, S. (1993). *The Psychology of Attitudes*. Harcourt brace Jovanovich college publishers.

Flake, J. K., & Fried, E. I. (2020). Measurement schmeasurement: Questionable measurement practices and how to avoid them. *Advances in Methods and Practices in Psychological Science, 3*, 456-465.

Flake, J. K., Pek, J., & Hehman, E. (2017). Construct validation in social and personality research: Current practice and recommendations. *Social Psychological and Personality Science, 8*, 370-378.

Flora, D. B. (2020). Your coefficient alpha is probably wrong, but which coefficient omega is right? A tutorial on using R to obtain better reliability estimates. *Advances in Methods and Practices in Psychological Science, 3*, 484-501.

Freelon, D. (2014). On the interpretation of digital trace data in communication and social computing research. *Journal of Broadcasting & Electronic Media, 58*, 59-75.

Hobbs, W. R., Burke, M., Christakis, N. A., & Fowler, J. H. (2016). Online social integration is associated with reduced mortality risk. *Proceedings of the National Academy of Sciences, 113*, 12980-12984.

Horn, J. L. (1965). A rationale and test for the number of factors in factor analysis. *Psychometrika*, *30*, 179-185.

Howe, L. C., & Krosnick, J. A. (2017). Attitude strength. *Annual Review of Psychology, 68*, 327-351.

Hutto, C., & Gilbert, E. (2014, May). Vader: A parsimonious rule-based model for sentiment analysis of social media text. In *Proceedings of the international AAAI conference on web and social media* (Vol. 8, No. 1, pp. 216-225).

Jorgensen, T. D., Pornprasertmanit, S., Schoemann, A. M., & Rosseel, Y. (2022). *semTools: Useful tools for structural equation modeling*. R package version 0.5-6. Retrieved from https://CRAN.R-project.org/package=semTools

Kaplan, R. M., & Saccuzzo, D. P. (2001). *Psychological testing: Principles, applications, and issues.* Wadsworth/Thomson Learning.

Kelley, T. L. (1927). *Interpretation of Educational Measurements.* World Book Co..

Kenny, D. A., Kaniskan, B., & McCoach, D. B. (2015). The performance of RMSEA in models with small degrees of freedom. *Sociological Methods & Research, 44*, 486-507.

Kitchin, R. (2014). Big Data, new epistemologies and paradigm shifts. *Big Data & Society*, *1*, 2053951714528481.

Lazer, D. M., Pentland, A., Watts, D. J., Aral, S., Athey, S., Contractor, N., ... & Wagner, C. (2020). Computational social science: Obstacles and opportunities. *Science*, *369*, 1060-1062.

Lazer, D., Hargittai, E., Freelon, D., Gonzalez-Bailon, S., Munger, K., Ognyanova, K., & Radford, J. (2021). Meaningful measures of human society in the twenty-first century. *Nature*, *595*, 189-196.

Lazer, D., Pentland, A., Adamic, L., Aral, S., Barabási, A. L., Brewer, D., ... & Van Alstyne, M. (2009). Computational social science. *Science*, *323*, 721-723.

Li, W., Aste, T., Caccioli, F., & Livan, G. (2019). Early coauthorship with top scientists predicts success in academic careers. *Nature Communications*, *10*, 5170.

Luttrell, A., & Sawicki, V. (2020). Attitude strength: Distinguishing predictors versus defining features. *Social and Personality Psychology Compass*, *14*(8), e12555.

McDonald R. P. (1999). *Test theory: A unified treatment*. Mahwah, NJ: Lawrence Erlbaum.

Mohammad, S., Kiritchenko, S., Sobhani, P., Zhu, X., & Cherry, C. (2016, June). Semeval-2016 task 6: Detecting stance in tweets. In *Proceedings of the 10th international workshop on semantic evaluation (SemEval-2016)* (pp. 31-41).

Ouyang, L., Wu, J., Jiang, X., Almeida, D., Wainwright, C., Mishkin, P., ... & Lowe, R. (2022). Training language models to follow instructions with human feedback. *Advances in Neural Information Processing Systems*, *35*, 27730-27744.

Park, M., Leahey, E., & Funk, R. J. (2023). Papers and patents are becoming less disruptive over time. *Nature, 613*, 138-144.

Petrocelli, J. V., Tormala, Z. L., & Rucker, D. D. (2007). Unpacking attitude certainty: Attitude clarity and attitude correctness. *Journal of Personality and Social Psychology, 92*, 30.

Petty, R. E., & Krosnick, J. A. (Eds.). (1995). *Attitude strength: Antecedents and consequences*. Laurence Erlbaum Associated.

Qin, C., Zhang, A., Zhang, Z., Chen, J., Yasunaga, M., & Yang, D. (2023). Is ChatGPT a general-purpose natural language processing task solver? *arXiv preprint arXiv:2302.06476.*

Revelle, W. (2023). *psych: Procedures for Psychological, Psychometric, and Personality Research*. Northwestern University, Evanston, Illinois. R package version 2.3.3, https://CRAN.R-project.org/package=psych.

Revelle, W., & Condon, D. M. (2019). Reliability from α to ω: A tutorial. *Psychological assessment, 31*, 1395.

Rios, K., DeMarree, K. G., & Statzer, J. (2014). Attitude certainty and conflict style: Divergent effects of correctness and clarity. Personality and Social Psychology Bulletin, 40, 819830.

Rosenbusch, H., Wanders, F., & Pit, I. L. (2020). The Semantic Scale Network: An online tool to detect semantic overlap of psychological scales and prevent scale redundancies. *Psychological Methods, 25*, 380–392

Rosseel Y. (2012). “lavaan: An R Package for Structural Equation Modeling.” *Journal of Statistical Software*, *48*, 1–36.

Roy, S., Nakshatri, N. S., & Goldwasser, D. (2022, November). Towards Few-Shot Identification of Morality Frames using In-Context Learning. In *Proceedings of the Fifth Workshop on Natural Language Processing and Computational Social Science (NLP+ CSS)* (pp. 183-196).

Simmons, J. P., Nelson, L. D., & Simonsohn, U. (2011). False-positive psychology: Undisclosed flexibility in data collection and analysis allows presenting anything as significant. *Psychological Science*, *22*, 1359-1366.

Skitka, L. (n.d.) *Measuring moral conviction*. Retrieved Feb 25, 2024 from https://lskitka.people.uic.edu/styled-7/styled-15/styled-8/index.html

Skitka, L. J. (2010). The psychology of moral conviction. *Social and Personality Psychology Compass, 4*, 267281

Skitka, L. J., Bauman, C. W., & Sargis, E. G. (2005). Moral conviction: Another contributor to attitude strength or something more? Journal of Personality and Social Psychology, 88, 895

Skitka, L. J., Hanson, B. E., Morgan, G. S., & Wisneski, D. C. (2021). The psychology of moral conviction. *Annual Review of Psychology, 72*, 347-366.

Tufekci, Z. (2014, May). Big questions for social media big data: Representativeness, validity and other methodological pitfalls. In *Proceedings of the International AAAI Conference on Web and Social Media* (Vol. 8, No. 1, pp. 505-514).

Visser, P. S., Bizer, G. Y., & Krosnick, J. A. (2006). Exploring the latent structure of strength-related attitude attributes. Advances in experimental social psychology, 38, 1-67.

Vosoughi, S., Roy, D., & Aral, S. (2018). The spread of true and false news online. S*cience*, *359*, 1146-1151.

Wagner, C., Strohmaier, M., Olteanu, A., Kıcıman, E., Contractor, N., & Eliassi-Rad, T. (2021). Measuring algorithmically infused societies. *Nature*, *595*, 197-204.

Weijters, B., & Baumgartner, H. (2012). Misresponse to reversed and negated items in surveys: A review. *Journal of Marketing Research, 49*, 737-747.

Yang, T., Shi, T., Wan, F., Quan, X., Wang, Q., Wu, B., & Wu, J. (2023, December). PsyCoT: Psychological Questionnaire as Powerful Chain-of-Thought for Personality Detection. In *Findings of the Association for Computational Linguistics: EMNLP 2023* (pp. 3305-3320).

Yang, Y., & Green, S. B. (2011). Coefficient alpha: A reliability coefficient for the 21st century?. *Journal of Psychoeducational Assessment*, *29*, 377-392.

Zhang, W., Deng, Y., Liu, B., Pan, S. J., & Bing, L. (2023). Sentiment analysis in the era of large language models: A reality check. *arXiv preprint arXiv:2305.15005*.

Ziems, C., Held, W., Shaikh, O, Chen, J., Zhang, Z. & Yang, D. (2024). Can large language models transform computational social science? *Computational Linguistics, 50,* 1-55

Zuwerink, J. R., & Devine, P. G. (1996). Attitude importance and resistance to persuasion: It's not just the thought that counts. Journal of Personality and Social Psychology, 70, 931.

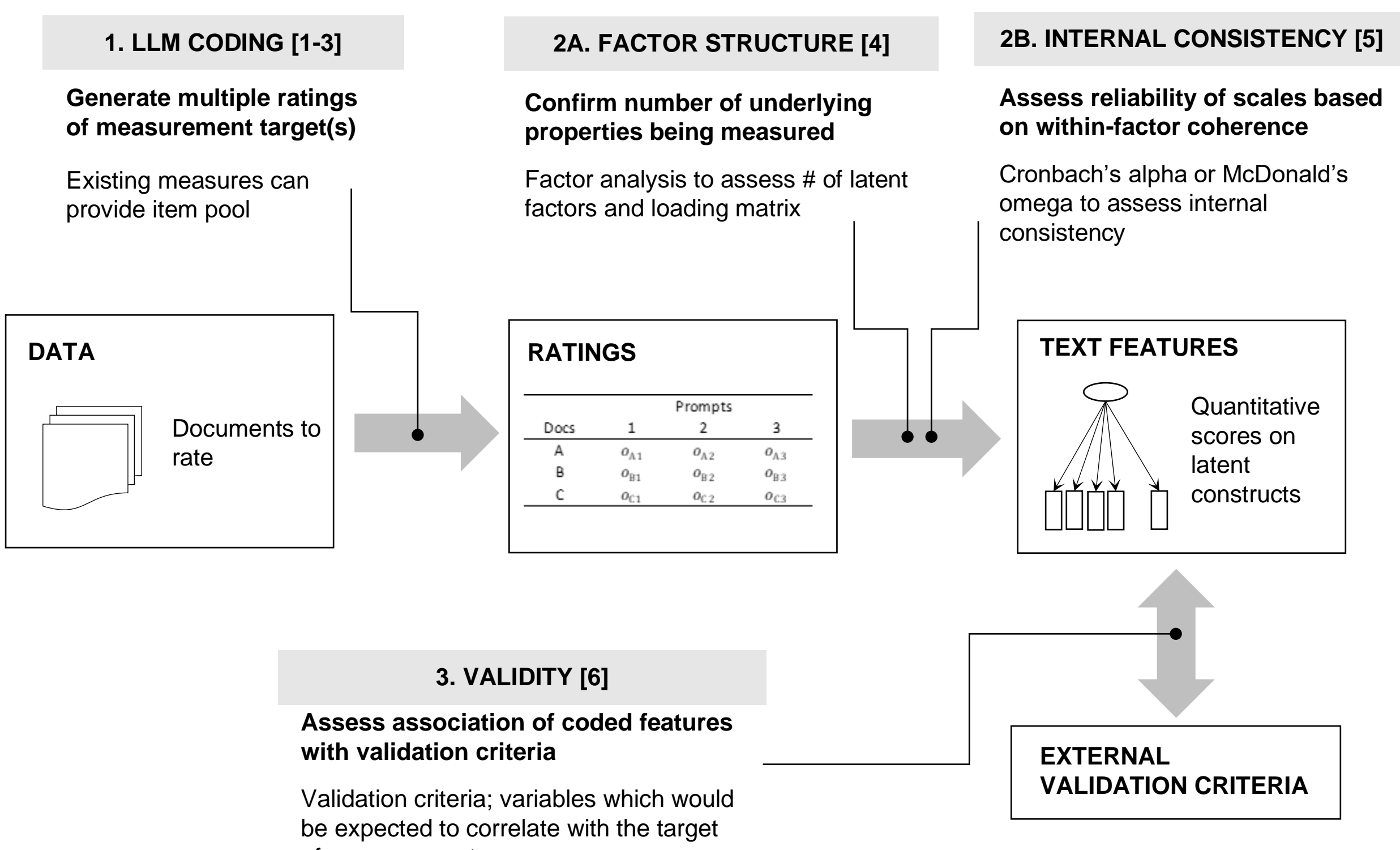

**Figure 1. Steps to score text features using large language models.** Variant prompts are used to generate multiple indicators of each proposed latent variable. The reliability of these measures is then assessed by (a) confirming the factor structure of these indicators and then (b) quantifying their internal consistency. Following this, validity is checked against external criteria. [Numbers in square brackets] refer to the associated steps in the workflow.

***Read the following text, then respond to the statement below it:***

***START OF TEXT***

**Title**

**Document**

***END OF TEXT***

***Based on this text, how much would you agree with the following statement:***

**Response item**

***Respond with one of the following items:***
***[strongly disagree, disagree, undecided, agree, strongly agree]***

[1] Initial instructions provide an overview of the task

[2] Text to be rated is added to prompt with clear start and end delimiter

[3] Instructions provide guidance on how to use response item

[4] Response item is added in the form of a statement (to match instructions)

[5] Explicit response options are provided

**Figure 2. Rating prompt template developed for attitude rating.** Template provides instructions to the model (1,3), fixed response options which can be converted to numerical ratings (5) and points to pipe in documents to be rated (2) and response items to rate them on (4; see Table 1 for examples).

**Table 1.** Response items for attitude property ratings. Items with an asterisk (*) are reverse coded.

| | | |
|---|---|---|
| **Attitude certainty** | Clarity_1 | The author is certain that they know what their true attitude on this topic really is |
| | Clarity_2 | The author is certain that the attitude they expressed towards this topic really reflects their true thoughts and feelings |
| | Clarity_3 | The author's true attitude towards the issue is clear in their mind |
| | Clarity_4 | The author is certain that the attitude they expressed towards this issue is really the attitude they have |
| | Correctness_1 | The author is certain that their attitude toward this issue is the correct attitude to have |
| | Correctness_2 | The author thinks other people should have the same attitude as them on this issue |
| | Correctness_3 | The author is certain that of all the possible attitudes one might have towards the issue, their attitude reflects the right way to think and feel about the issue |
| **Attitude Importance** | Importance_1 | The author's attitude towards this issue is very important to them personally |
| | Importance_2* | The author does not care personally about this issue |
| | Importance_3* | The author does not have very intense feelings about this issue |
| | Importance_4 | The author is personally very concerned about this issue |
| | Importance_5 | This issue is important to the author personally |
| | Importance_6 | The author personally cares about this issue |
| | Importance_7 | This issue means a lot to the author |
| | Importance_8 | This issue is important to the author as compared to other issues |
| **Moral conviction** | Moral_1 | The author's position on this issue is a reflection of their core moral beliefs and convictions |
| | Moral_2 | The author's position on this issue is connected to their beliefs about fundamental right and wrong |
| | Moral_3 | The author's position on this issue is based on moral principle |
| | Moral_4 | The author's position on this issue is a moral stance |

**Table 2.** Factor structure and internal consistency of the LLM-based measurement. Left-hand columns show results from an exploratory factor analysis (to identify factor structure and identify well-performing items). Right-hand columns show loadings and reliability measures from confirmatory factor models fitted to a separate confirmatory dataset.

| | | Exploratory factor model (promax rotation) | | | | | Confirmatory factor models (unifactor) | | |
|---|---|---|---|---|---|---|---|---|---|
| | | F1 | F2 | F3 | F4 | F5 | Cert. | Imp. | Moral. |
| Factor loadings | Clarity_1 | 0.34 | | 0.72 | | | | | |
| | Clarity_2 | 0.66 | | | | | 0.88 | | |
| | Clarity_3 | 0.55 | | -0.35 | | 0.31 | 0.60 | | |
| | Clarity_4 | 0.78 | | | | | 0.92 | | |
| | Correctness_1 | 0.32 | | 0.76 | | | | | |
| | Correctness_2 | 0.39 | | | | | 0.74 | | |
| | Correctness_3 | | | 0.82 | | | | | |
| | Importance_1 | | | | 0.56 | | | | |
| | Importance_2* | | | | 0.90 | | | | |
| | Importance_3* | | | | 0.77 | | | | |
| | Importance_4 | | 1.00 | | | | | 0.84 | |
| | Importance_5 | | 0.72 | | | | | 0.91 | |
| | Importance_6 | | 0.89 | | | | | 0.85 | |
| | Importance_7 | | 0.83 | | | | | 0.92 | |
| | Importance_8 | | 0.77 | | | | | 0.91 | |
| | Moral_1 | | | 0.54 | | 0.44 | | | 0.80 |
| | Moral_2 | | | | | 0.63 | | | |
| | Moral_3 | | | 0.90 | | | | | 0.98 |
| | Moral_4 | | | 0.96 | | | | | 0.94 |
| Robust CFI | | | | | | | 0.98 | 1.00 | NA |
| Robust TLI | | | | | | | 0.95 | 0.99 | NA |
| Robust RMSEA | | | | | | | 0.15 | 0.12 | NA |
| Cronbach's alpha | | | | | | | 0.87 | 0.82 | 0.83 |
| McDonald’s omega | | | | | | | 0.95 | 0.82 | 0.83 |

**Table 3.** Logistic regression of stance status (3a favour vs. none; 3b oppose vs. none) on LLM-coded attitude property scores. 'Control' models include message sentiment. 'Comparison prompt' models measure attitude strength using a single face-valid LLM prompt.

| a. | Certainty | | Importance | | Moral conviction | | Comparison prompt | |
|---|---|---|---|---|---|---|---|---|
| | Base | Control | Base | Control | Base | Control | Base | Control |
| | | | | *Model coefficients* | | | | |
| Intercept | -0.74 *** | -0.76 *** | -1.41 *** | -1.44 *** | -0.52 *** | -0.53 *** | -0.63 * | -0.63 * |
| Certainty | 0.78 *** | 0.79 *** | | | | | | |
| Importance | | | 1.22 *** | 1.24 *** | | | | |
| Moral conviction | | | | | 0.70 *** | 0.70 *** | | |
| Comparison prompt | | | | | | | 0.44 * | 0.44 * |
| Sentiment | | 0.15 | | 0.17 | | 0.13 | | 0.06 |
| | | | | *Model fit* | | | | |
| AIC | 897.58 | 898.87 | 871.09 | 872.13 | 884.58 | 885.98 | 926.25 | 928.12 |
| Pseudo-$R^2$ | 0.04 | 0.04 | 0.07 | 0.07 | 0.05 | 0.05 | 0.01 | 0.01 |

| b. | Certainty | | Importance | | Moral conviction | | Comparison prompt | |
|---|---|---|---|---|---|---|---|---|
| | Base | Control | Base | Control | Base | Control | Base | Control |
| | | | | *Model coefficients* | | | | |
| Intercept | 0.74 *** | 0.75 *** | 0.79 *** | 0.79 *** | 0.76 *** | 0.77 *** | -0.37 | -0.36 |
| Certainty | -0.05 | -0.05 | | | | | | |
| Importance | | | -0.09 | -0.09 | | | | |
| Moral conviction | | | | | -0.16 ** | -0.16 ** | | |
| Comparison prompt | | | | | | | 0.75 *** | 0.75 *** |
| Sentiment | | -0.16 | | -0.16 | | -0.15 | | -0.15 |
| | | | | *Model fit* | | | | |
| AIC | 1297.73 | 1298.39 | 1296.78 | 1297.46 | 1289.85 | 1290.65 | 1276.32 | 1277.28 |
| Pseudo-$R^2$ | 0.00 | 0.00 | 0.00 | 0.00 | 0.01 | 0.01 | 0.02 | 0.02 |

NB *** $p < .001$; ** $p < .01$; * $p < .05$; Pseudo-$R^2$ calculated using the McFadden method